\def\ket{\vert \vert  \{ \emptyset \} \rangle}
  \def\ket2{\vert \vert \otimes \{ R \} \rangle}

\def\.#1{\mathaccent 95#1}
\def\^#1{\mathaccent 94 #1}
\def\~#1{\mathaccent "7E #1}

\def\eq{\enskip =\enskip}
\def\pls{\enskip +\enskip}
\def\mns{\enskip -\enskip}

\def\trans{{\cal T}}
\def\proj{{\cal P}}

\def\P{$P$}

  \def\proj{{\cal P}}
  \def\trans{{\cal T}}
  \def\ket{\vert \vert  \{ \emptyset \} \rangle}
  \def\ket2{\vert \vert \otimes \{ R \} \rangle}

\def\d{{\cal D}}

\documentstyle[12pt]{ioplppt}
\begin{document}
\title{ Magnetism and magnetic asphericity in  NiFe alloys}
\author{Biplab Sanyal\footnote{Corresponding author. Fax : 0091 333343477,
 Phone : 0091 333583061, E-mail : biplab@boson.bose.res.in}
 and Abhijit Mookerjee}
\address{S.N.Bose National Centre for Basic Sciences, JD Block,
Sector 3, Salt Lake City, Calcutta 700091, India}

\begin{abstract}
We here study magnetic properties of Ni$_{x}$Fe$_{1-x}$ using
Augmented space recursion technique coupled with tight-binding
linearized muffin tin orbital method. Also the spectral properties
of this alloy has been studied here.
\end{abstract}
{\parindent 0pt
{\bf Keywords :} Magnetism, alloys, augmented space recursion}
\baselineskip 25pt

\section{Introduction}

NiFe alloys have been extensively studied earlier. Experimental data
have been available for decades. Neutron scattering has probed the local
magnetic moments \cite{kn:sw,kn:col,kn:nishi,kn:br,kn:ito}. Theoretical approaches include the
Hartree-Fock approaches of Hasegawa and Kanamori \cite{kn:hk} and
Kanamori \cite{kn:k}. These authors assumed the density of states to be
steeple models, and were therefore only qualitative. They used the
coherent potential approach (CPA) of Velick\'y
\etal \cite{kn:v} in order to describe disorder and configuration averaging
. Mishra and Mookerjee \cite{kn:mm} have used the Hartree-Fock model
of Hasegawa and Kanmori but the averaged density of states was obtained
from the augmented space technique based
cluster CPA. Podg\'orny \cite{kn:pd} studied NiFe using the supercell linearized
muffin-tin orbitals method (LMTO) while Lipi\'nski \cite{kn:l} applied
the tight-binding LMTO-CPA.

NiFe alloys have interesting behaviour. The magnetic behaviour of Ni is
anomalous in the sense that the local magnetic moment of Ni depends
sensitively on its immediate surrounding. The local magnetic moment on
a Ni atom depends upon the number of Ni neighbours it has. The CPA which
replaces the random neighbourhood of an atom in a solid solution by an effective,
averaged medium cannot reflect this behaviour. The aim of this communication
is to study NiFe by the augmented space recursion (ASR) technique \cite{kn:asr}.
Recently we have shown that the ASR allows us to go beyond the CPA without
violating the essential herglotz analytic properties and take into account
the effect of local neighbourhood \cite{kn:sdm}. The ASR will be based on
the TB-LMTO hamiltonian

\begin{eqnarray}
H \eq \sum_{RL} \^C_{RL} \proj_{RL} \pls \sum_{RL}\sum_{R'L'} \^\Delta^{1/2}_{RL}
S_{RL,R'L'}\^\Delta^{1/2}_{R'L'} \trans_{RL,R'L'} \nonumber \\
\^C_{RL} \eq C^{B}_{RL} \pls \left(C^{A}_{RL}-C^{B}_{RL}\right)\;n_{R} \nonumber\\
\^\Delta^{1/2}_{RL} \eq \Delta^{B1/2}_{RL}\pls \left(\Delta^{A1/2}_{RL} - \Delta^{B1/2}_{RL}
\right)\; n_{R}
\end{eqnarray}

Here $\proj_{RL}$ and $\trans_{RL,R'L'}$ are projection and transfer operators in the hilbert
space spanned by the tight binding basis $\vert RL\rangle$ and $n_{R}$ is a random occupation
variable which is 1 if the site $R$ is occupied by an atom of the A type and 0 if not. The augmented
space hamiltonian replaces the random occupation variable by operators $M_{R}$ of rank 2. For
models without any short-range order

\[ M_{R} \eq x \proj_{\uparrow}^{R} \pls (1-x) \proj_{\downarrow}^{R} \pls \sqrt{x(1-x)}
\left( \trans_{\uparrow\downarrow}^{R}+\trans_{\downarrow\uparrow}^{R} \right) \]

\[ \vert \uparrow \rangle \eq \left( \sqrt{x}\vert 0 \rangle \pls \sqrt{1-x}\vert 1 \rangle \right) \]
\[ \vert \downarrow \rangle \eq \left(\sqrt{1-x}\vert 0\rangle \mns \sqrt{x}\vert 1 \rangle \right) \]

The ASR was carried out with LDA self-consistency. The computations included scalar relatavistic corrections.
The exchange potential of von Barth-Hedin was used. We used the choice of flexible Wigner Seitz radii
for Ni and Fe as suggested by Kudrnovsk\'y and Drchal \cite{kn:kd} in order to ensure the atomic spheres
to be neutral and avoid the calculation of the Madelung energy.

The  recursion method then expresses the Green functions as continued fraction expansions. The continued
fraction coefficients are exactly obtained upto eight levels and the terminator suggested by Luchini
and Nex \cite{kn:ln} is used to approximate the asymptotic part. The convergence of this procedure has
been discussed by Ghosh \etal \cite{kn:gdm}. The local charge densities are given by :

\begin{equation}
\rho^{\lambda}_{\sigma}(r) \eq (-1/\pi) \Im m \sum_{L} \int_{-\infty}^{E_{F}} dE \ll G_{LL}^{\lambda,\sigma}(r,r,E)\gg
\end{equation}

Here $\lambda$ is either $A$ or $B$. The local magnetic moment is

\[ m^{\lambda} \eq \int_{r<R_{WS}} d^{3}r\; \left[\rho_{\uparrow}(r)\mns \rho_{\downarrow}(r)\right] \]

 We have also obtained the spectral densities and complex band structures for 50-50 Ni-Fe
using the ASR in k-space as suggested by Biswas \etal \cite{kn:ksp}.

For high concentrations of Ni the
alloy forms a face centred cubic solid solution. With decreasing Ni concentration,
at the invar concentration of about 30$\%$, the system undergoes a
structural phase transition to a body centred cubic solid solution. At this
transition there is a sharp decrease of Wigner-Seitz radius. This leads
to larger overlap of the $d$-bands, and hence, using the Stoner criterion, a
decrease of magnetization. We shall limit ourselves to the face centred cubic
region beyond the invar concentration.

\section{Results and Discussion}

Figures 1(a) and (b)  show the Ni and Fe partial density of states for Ni concentrations varying between
90 and 40 $\%$. We note that the Ni densities hardly change with concentration, while the Fe densities
change considerably.  For the minority spin partial densities on Fe, the structures below the Fermi
energy do not change with concentration. The peak at around 0.0 Ryd grows with
increasing Ni concentration. However, this structure is above the Fermi level and does not contribute to
the magnetic moment. For the majority spin partial
densities on Fe, the peak around -0.4 Ryd grows with increasing Ni concentration
. This ensures that the Ni local magnetic moment remains almost concentration
independent, while the Fe local	 moment increases with Ni concentration. Figure 2 shows the Ni and Fe
moments as well as the average moment as functions of Ni concentration. The results
 agree rather well with earlier CPA results. In addition we have shown the experimental results on
the average magnetic moment by Crangle and Halam \cite{kn:crang} as well as the experimental data on
local magnetic moments on Fe and Ni by Shull and Wilkinson \cite{kn:sw}, Collins \etal \cite{kn:col}
and Nishi \etal \cite{kn:nishi}. The experimental data are reasonably reproduced, as well as CPA did earlier.

  Both the CPA and our results indicate a slight increase of the Fe  magnetic moment
as the Ni concentration increases. The experimental data with its large error bars tell us little about this
trend with certainty. The asphericity, which is measured by the ratio of the $t_{2g}$ and $e_{g}$
contributions to the magnetic moment, is shown in figure 3. The results indicate that the Ni
moment distribution is highly anisotropic. The $t_{2g}$ dominates the magnetization as well as
the density of states peaks near the Fermi level. We also compare our results with neutron
scattering experimental data of Brown \etal \cite{kn:br} and Ito \etal \cite{kn:ito}. The asphericity
trend with concentration is reproduced, but our average asphericities are slightly larger.
The asphericity decreases with decreasing Ni
concentration. The Fe moment distribution is almost spherical throughout the concentration range, with
a slight decrease with decreasing Ni concentration. The average asphericity agrees reasonably well
with experiment \cite{kn:br,kn:ito}.

Figures 4 (a) and (b)  gives the spectral densities for k-vectors going from the $\Gamma$ to the $X$ point
for the $e_{g}$ and $t_{2g}$ bands. The
results clearly show the splittng between the majority and minority spin bands. The imaginary part of
the self-energy which measures the disorder scattering life-time, is clearly k-dependent and
is more prominent at the $X$ point and least at the $\Gamma$ point. This is in contrast to the
CPA calculations where the life-times are almost k-independent. Angle-resolved photoemmision experiments
show the spectral behaviour of alloys and the nature of fuzzy fermi surfaces can be obtained
from Compton scattering and positron annihilation experiments. The complex bands are obtained
from the peaks and widths in the spectral densities. These are shown in figure 5 (a) and (b).
 We note that the d-bands of Fe and Ni overlap considerably, and although the
$e_{g}$ and $t_{2g}$ type bands are evident, they are broadened by disorder.
The lines shown in figure 5 are to guide the eye.
The statements made above based on the spectral functions are clearly seen in this figure. The
Fermi level crosses the minority spin bands. These bands have considerable life-times and the Fermi
surface should have this width associated with it. It should be interesting to look at the fuzzy
Fermi surface experimentally.

NiFe alloys tend to exhibit short-ranged order. The augmented space recursion is ideally suited to
take into account effects of short-ranged order. The formalism to include this has been introduced
earlier by Mookerjee and Prasad \cite{kn:sro} and applied to alloy systems by Saha \etal \cite {kn:sdm}.
We have carried out the calculation of the magnetic moment of the 50-50 NiFe alloy as a function
of the Warren-Cowley short-ranged order parameter $\alpha$. The magnetic moment of Fe is hardly affected
by short-ranged order. However, in the region where $\alpha < 0$ indicating ordering, Ni is more likely
to have Fe neighbours as compared to the case without short-ranged order. Here the magnetic moment on
Ni increases. Similarly, when $\alpha > 0$, Ni atoms segregate and are less likely to have Fe as their
neighbours and the Ni moment decreases.This is shown in figure 6.  We conclude that extra moment is induced on Ni by Fe atoms
in its vicinity. This induced moment on Ni is most sensitive to short-ranged order in the alloy.

\section*{Figure Captions}
\begin{description}
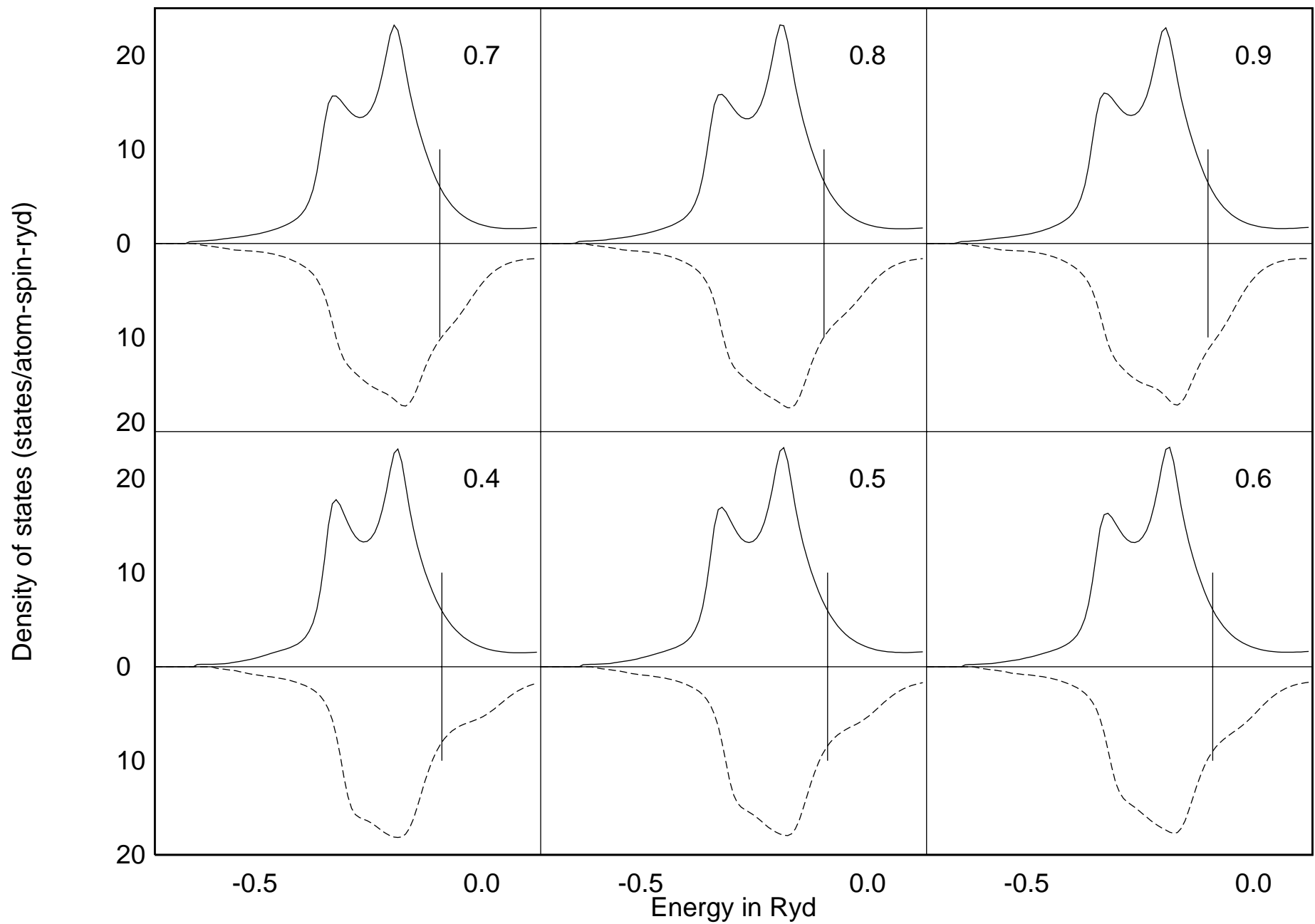
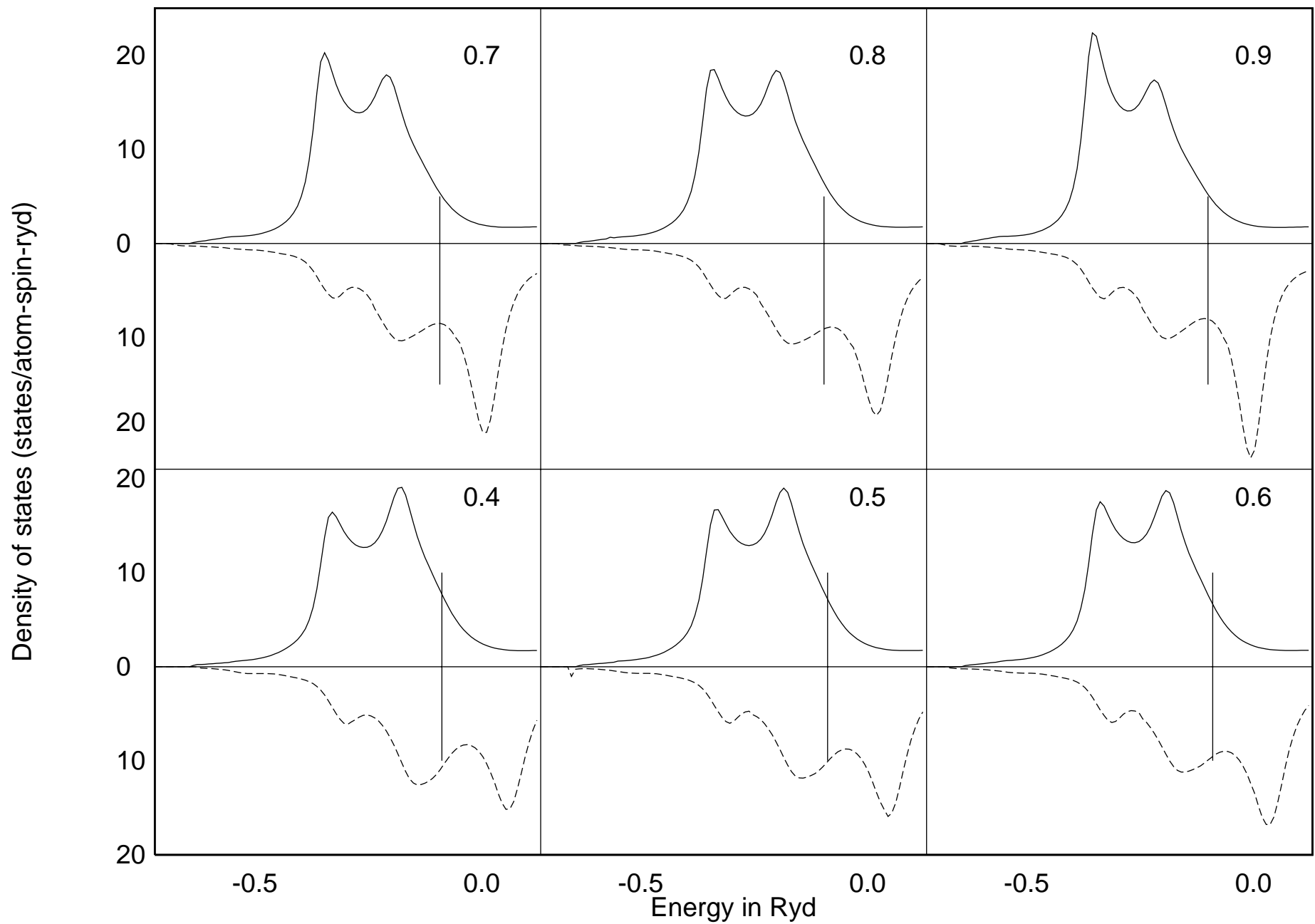
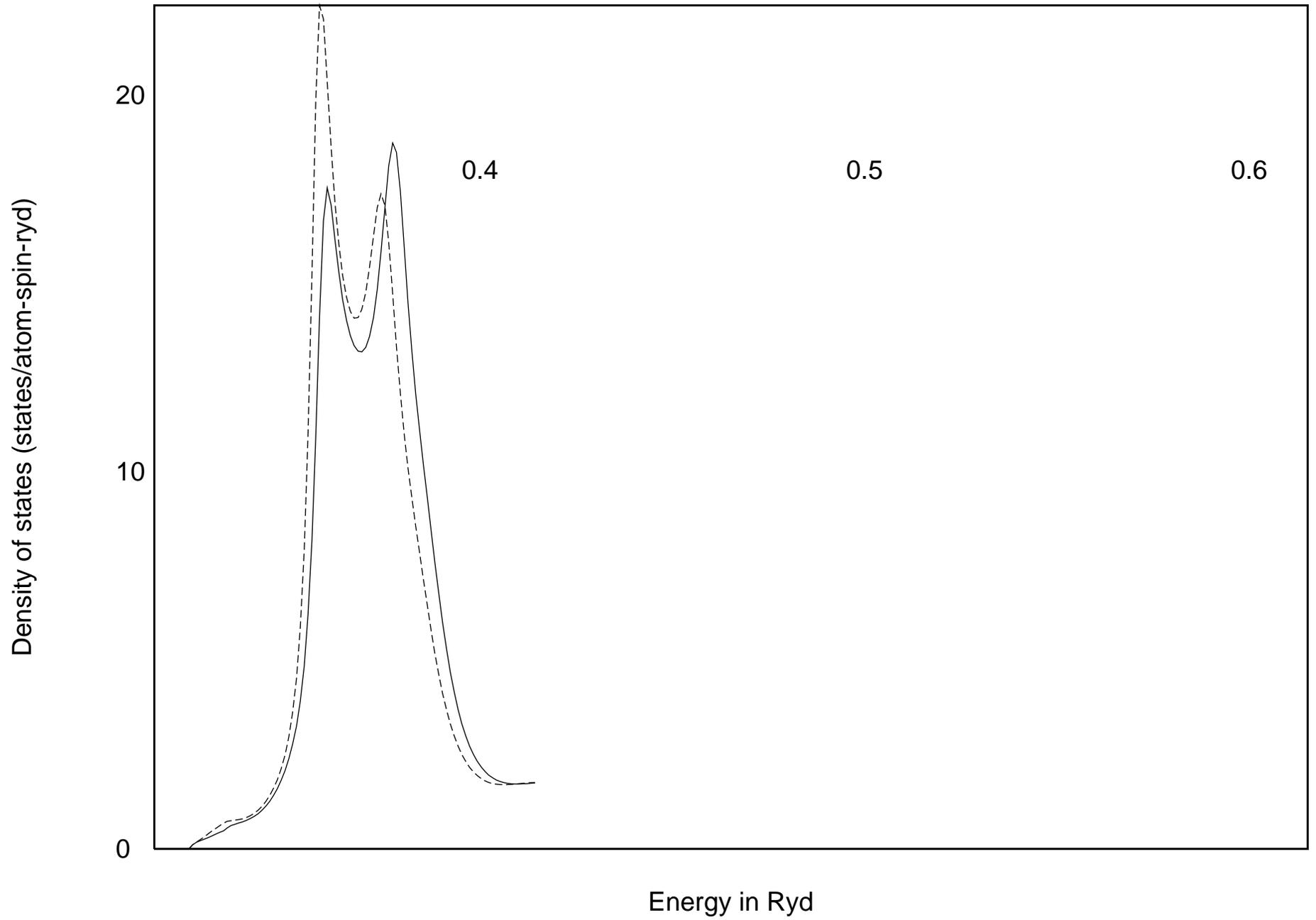
\item[Figure 1](a) Partial densities of states at Ni sites.
(b) Partial densities of states at Fe sites.
 The concentrations of Ni are shown as are
the Fermi energies (vertical lines).
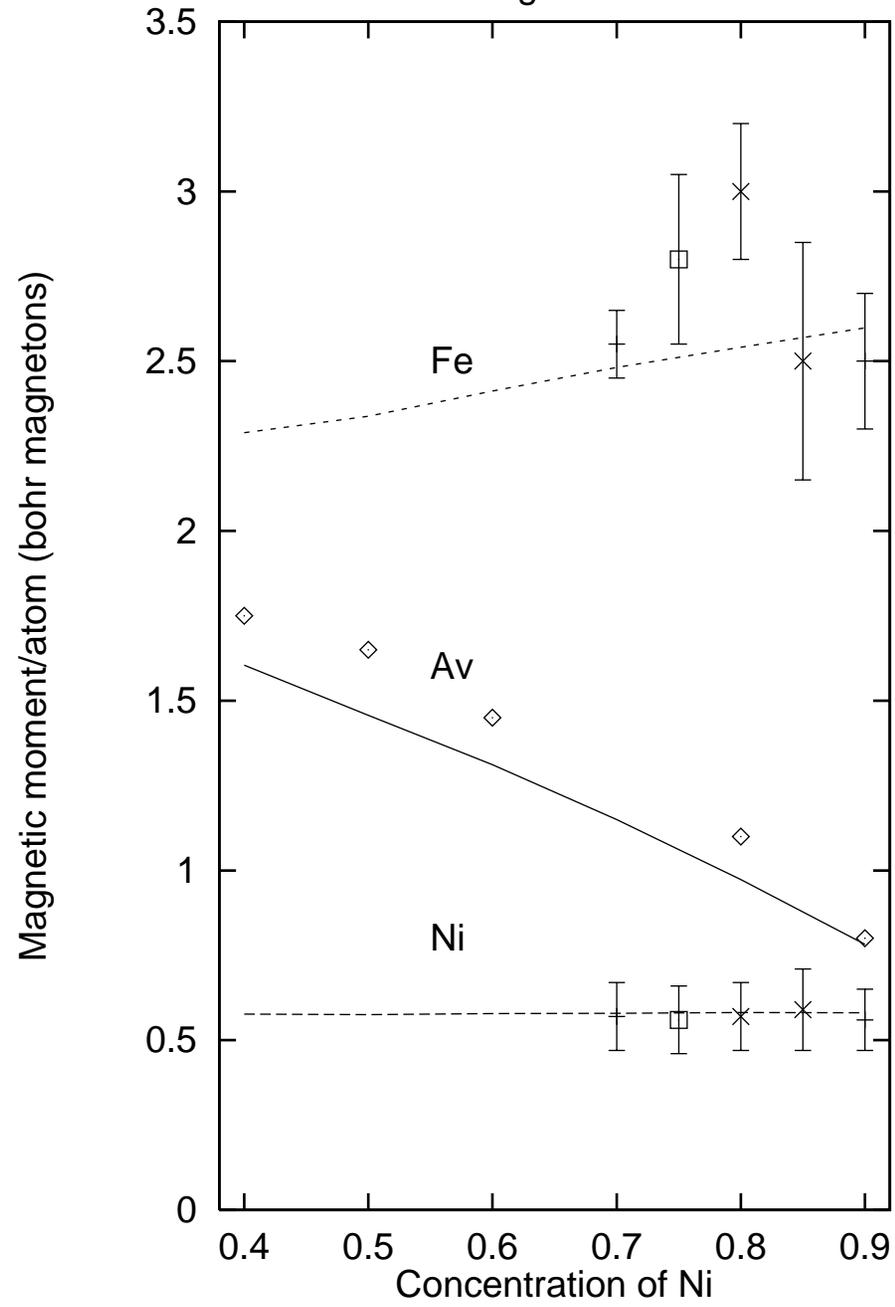
\item[Figure 2] Magnetic moments on Fe and Ni sites and the averaged magnetic moment as functions of
Ni concentration. The diamond marks indicate the experimental results of
Crangle and Halam, the dashed points of Shull and Wilkinson, the square points of Collins and Lowde
and the crossed points of Nishi \etal.
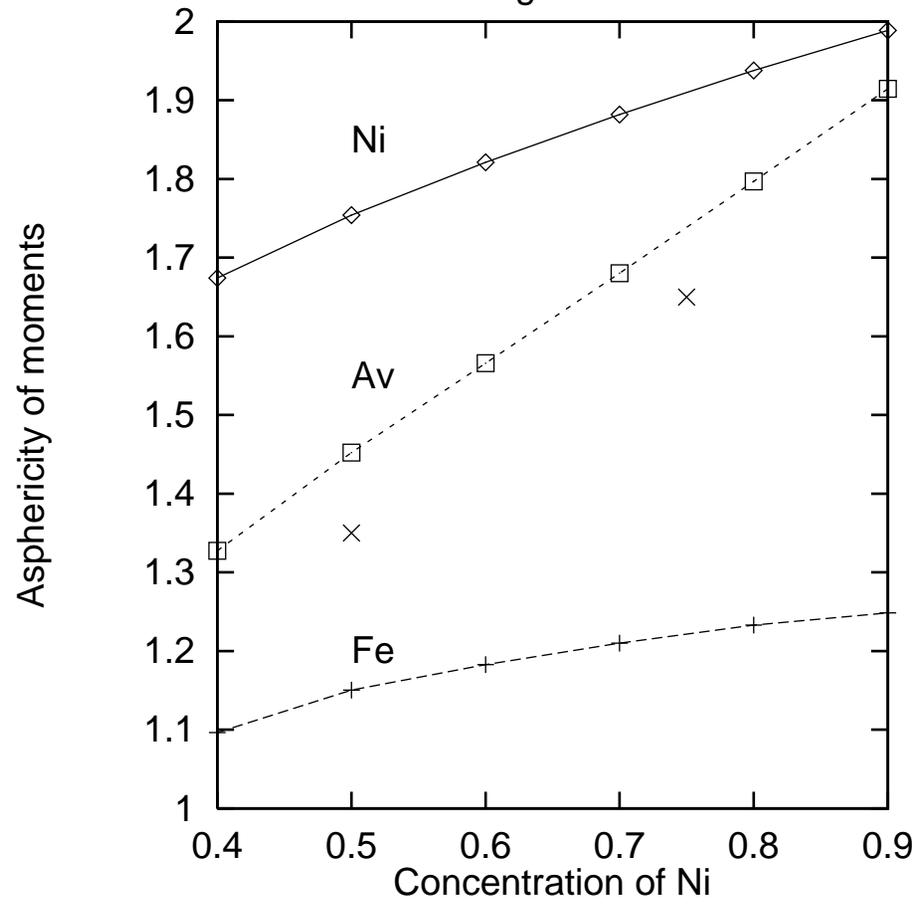
\item[Figure 3] Asphericity of moment distribution on Fe and Ni sites
as functions of Ni concentration. The cross marks indicate the neutron
scattering results of  Brown \etal and Ito \etal.
\item[Figure 4] The spectral densities for (a) $e_{g}$ and (b) $t_{2g}$ states along the $\Gamma$ to $X$
direction for 50-50 FeNi. The bold curves are for the majority spins and
 the dotted curve for the minority spins. For both (a) and (b), the wave
vectors of figures are (from top to bottom) (1,0,0), (0.75,0,0),(0.5,0,0),(0.25,0,0) and (0,0,0)
in units of 2$\pi$/a, where a is the lattice constant.
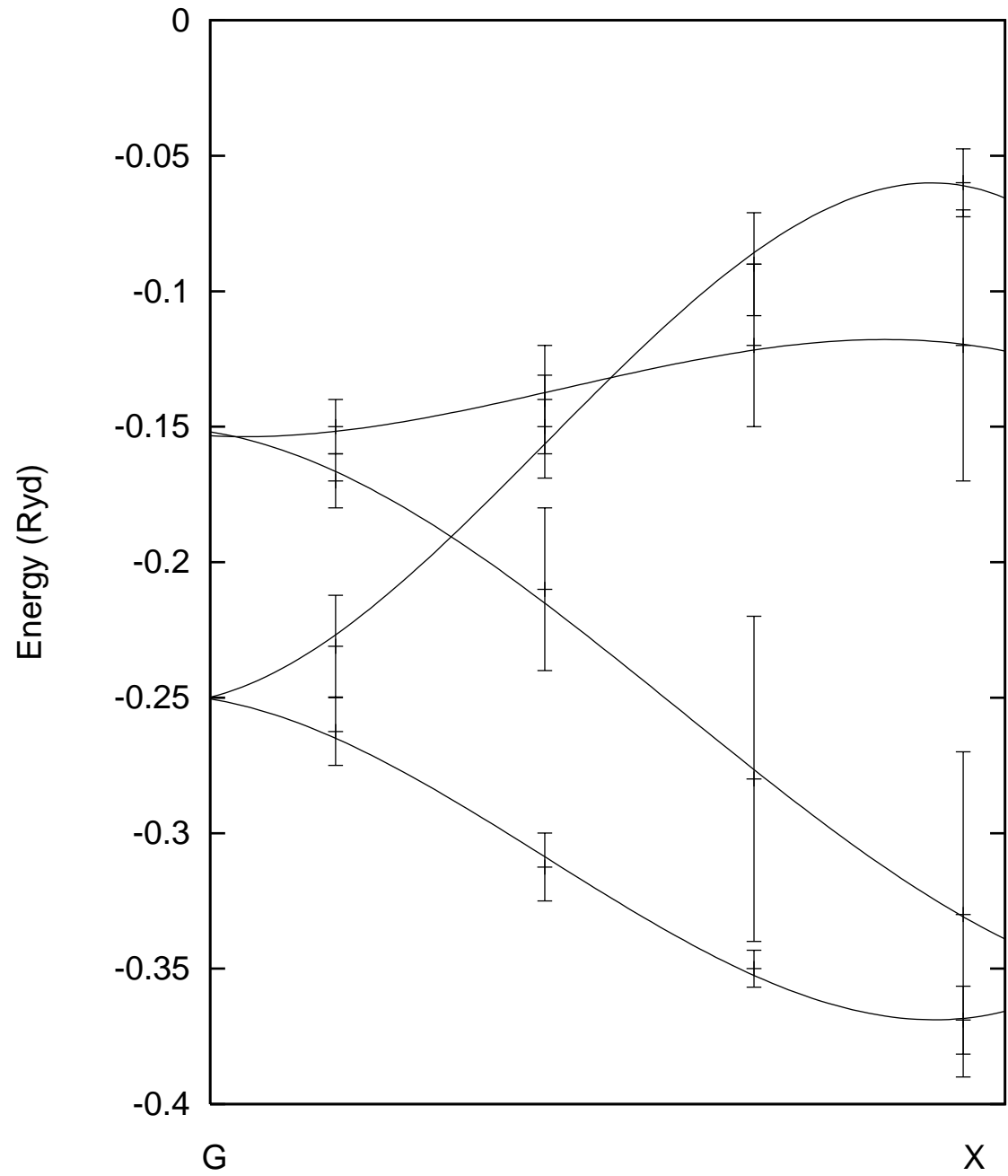
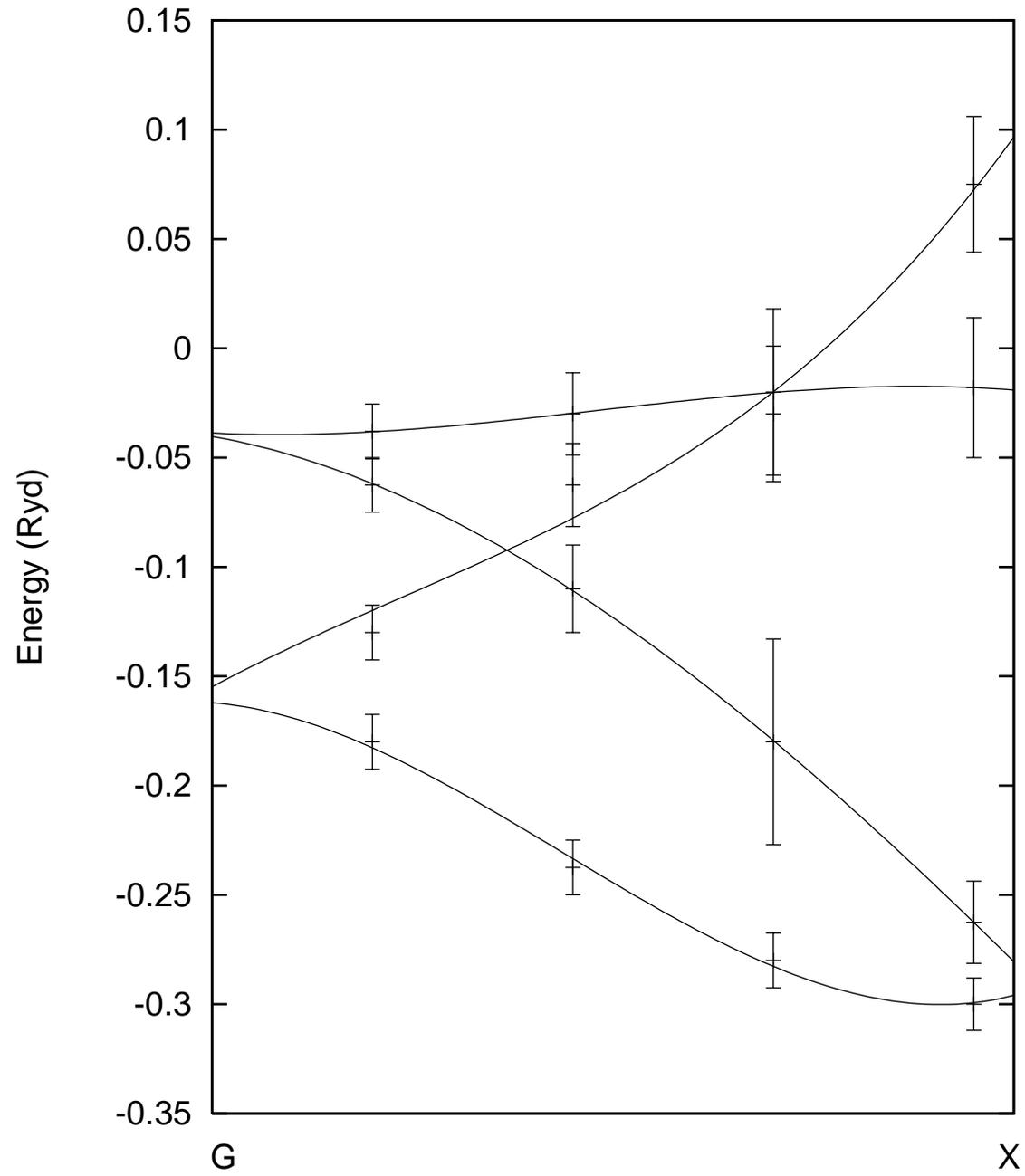
\item[Figure 5] Complex band structure for $d$-bands of 50-50 FeNi (a) For the majority
spin bands (b) for the minority spin bands. The widths of the bands are marked and
the lines are only to guide the eye.
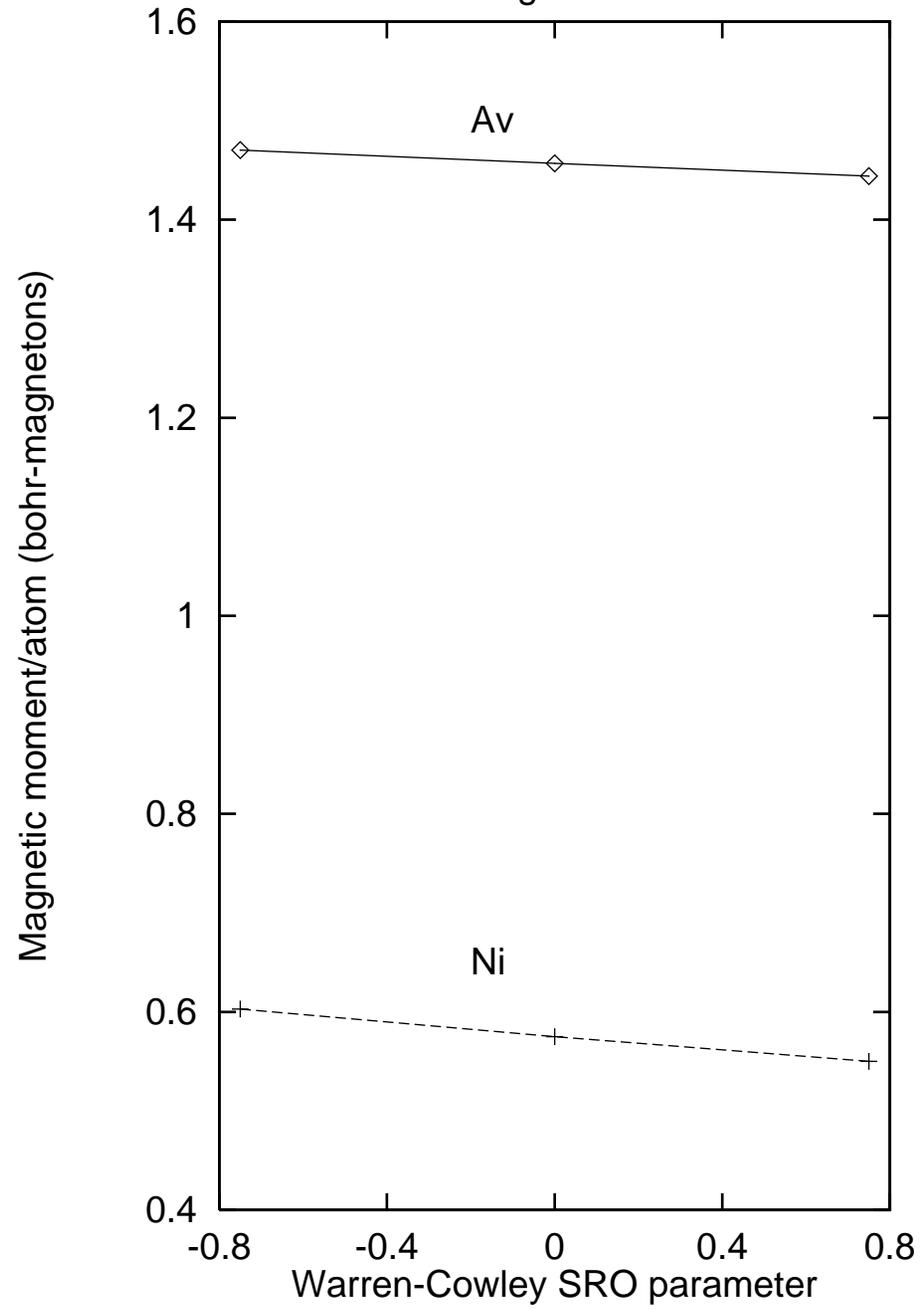
\item[Figure 6] The averaged magnetic moment and the magnetic moment of Ni as functions of
the Warren-Cowley short-ranged order parameter for 50-50 NiFe.
\end{description}
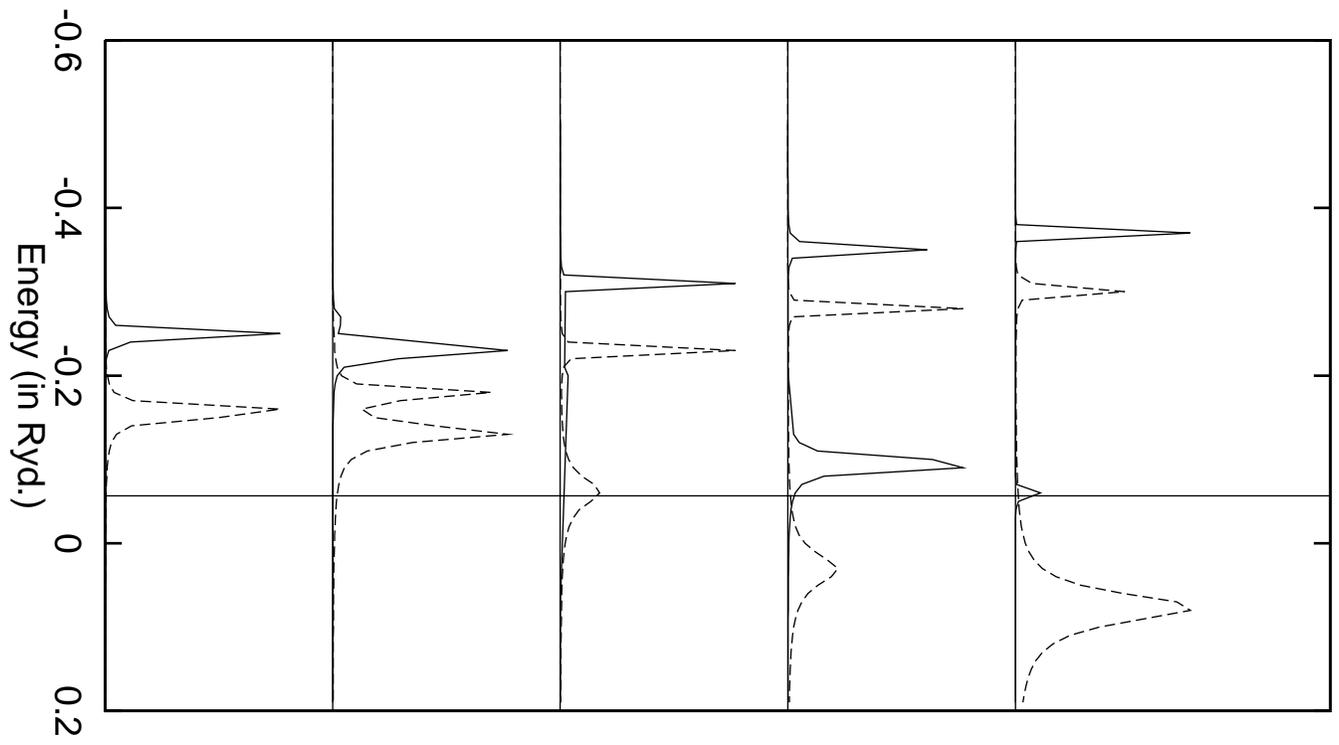
\end{document}